\documentclass[prl,aps]{revtex4}
\usepackage{graphicx}

\begin{document}

\title{Bessel-X waves: superluminal propagation and the Minkowski space-time}
\author{D. Mugnai}
\affiliation{``Nello Carrara'' Institute of Applied Physics, CNR
Florence Research Area \\ Via Madonna del Piano 10, 50019 Sesto
Fiorentino, Italy }

\begin{abstract}
\vspace{1 cm}
 Superluminal behavior has been extensively studied
in recent years,  especially with regard to the topic of
superluminality  in the propagation of a signal. Particular
interest has been devoted to Bessel-X waves propagation, since
some experimental results showed that these waves have both phase
and group velocities greater that light velocity c. However,
because of the lack of an exact definition of signal velocity,  no
definite answer about the signal propagation (or velocity of
information) has been found. The present paper is a short note
that deals in a general way with this vexed question. By analyzing
the field of existence of the Bessel X-wave in pseudo-Euclidean
space-time, it is possible to give a general description of the
propagation, and to overcome the specific question related to a
definition of signal velocity.
\end{abstract}

\maketitle

The propagation of a Bessel-X wave (or Bessel beam) is of great
interest in physics. This interest is due to the unusual features
which characterize these waves: they are non-diffracting
\cite{dur87,dur87-1,zio,spr,dur91,tan,rei}, and show superluminal
behaviour in both phase and group velocities
\cite{saa,mug00,ale,bes,zam,saa1}. Because of these two
characteristics, the interest that has  arisen with regard to
Bessel-X waves has led to several theoretical and experimental
investigations, none of which has as yet been able to provide
definite information on signal velocity.

For propagation in vacuum (or air),  the scalar field of a
Bessel-X wave propagating along the $z$-axis of a cylindrical
coordinate system ($\rho ,\:\psi ,\:z$) is given by:
\begin{equation} 
u(\rho ,\:\psi ,\:z)=AJ_0(k_0\rho\sin\theta_0) \exp(ik_0z\cos
\theta_0)  \exp(-i\omega t) \:,
 \label{u}
\end{equation}
where $A$ is an amplitude factor, $\theta_0$ is the parameter
which characterizes the aperture of the beam (Axicon angle), and
$k_0$ is the wavenumber in the vacuum. Function $J_0$ denotes the
zero-order Bessel function of first kind, which, apart from
inessential factors, can be written as \cite{abr}
\begin{equation}
J_0(x)=\int_0^{\pi} \exp (ix\cos\varphi )\,d\varphi .
 \label{delta}
 \end{equation}
The beam is rotationally symmetric and is thus independent of the
angular coordinate $\psi$. Equation  (\ref{u})
 is obtained from
the superposition of infinite plane waves of the same amplitude,
each propagating in a different direction and forming the same
angle $\theta_0$ with the $z$-axis.

In looking at Eq. (\ref{u}), we note that the dependence on $t$
and $z$ occurs only through the quantity
\begin{equation}
\frac{z}{c}\, \cos\theta_0 -t
 \label{tau}
\end{equation}
and, therefore, the beam (\ref{u}) takes the same value after a
time, $dt$, equal to $(dz/c)\cos\theta_0$: that is, it propagates
with a velocity
\begin{equation}
v= \frac{c}{\cos\theta_0}\:, \label{vsig}
\end{equation}
greater than $c$. It is well-known that, in the presence of
anomalous dispersion, group velocity can largely overcome the
speed of light in vacuum, and can even become negative. In the
absence of dispersion, the situation is different, since the three
velocities which characterize the propagation (phase, group, and
signal velocities) tend to coincide. This statement is valid not
only for waves, but also for  pulses, since all components at
different frequencies propagate with the same velocity. In fact,
in the absence of dispersion the pulse does not suffer
``reshaping'', a process that is always present in dispersive
systems. Moreover, if we insert a spectral function (depending
only on the frequency) in (\ref{u}), this presence does not modify
the situation and the previous conclusion is found to be true also
for a pulse.

\begin{figure}[t]
\includegraphics[width=0.6\textwidth]{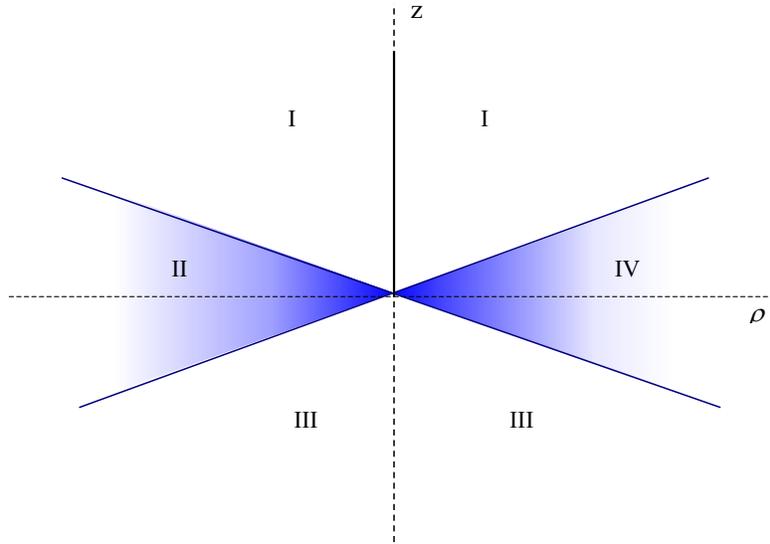}
 \caption{ Representation of zones of existence of a Bessel pulse
in the $\rho ,z$-plane, for $t=0$ (for $t> 0$, the diagram moves
with velocity $v = c/\cos\theta_0$ in the direction of $z$-axis).
In zones I and III the field is zero, while in II and IV its value
is different from zero: that is, the field exists only inside the
double surface delimited by the straight lines $\rho
=|z\cot\theta_0-(c/\sin\theta_0)t|$. For $\rho =0$, the field
reduces to a $\delta$ pulse. Zone II has no physical interest
since $\rho \geq 0$ .}
 \label{cono}
\end{figure}
Let us now analyze the propagation of a signal.

According to Brillouin \cite{bri},  a signal can be defined as a
pulse of finite temporal extension, that is, of an infinite
extension in the frequency domain. Thus, we can ``construct"  a
signal $U(\rho ,z,t)$ by superimposing a set of Bessel beams which
differ by the value of the frequency $\omega$ and have the same
value of the parameter $\theta_0$, as well as the same amplitude
and phase at $t=0$, that is:

\begin{figure}
\includegraphics[width=0.5\textwidth]{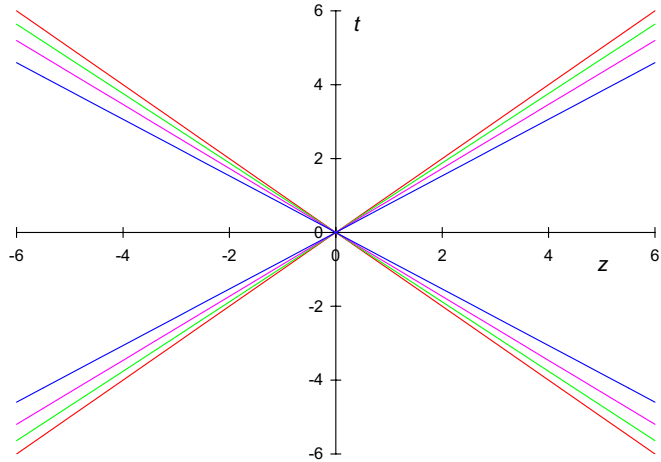}
  \caption{{\small Bessel beam velocities for
three different values of parameter $\theta_0$, in the $z-t$
plane, for $\rho=0$ (Euclidean space) . The red line indicates
light velocity $c$ (for $\theta_0=0$) taken here as equal to 1.
Green, pink, and blue lines represent the beam velocities for
$\theta_0 = 20^\circ,\, 30^\circ$, and $4^0\circ$, respectively.
For $\rho\neq 0$, the intersection point changes its position with
no modification in the line behavior.}} \label{beam-velocity}
\end{figure}
\begin{equation}
U(\rho ,z,t)= \int_{-\infty}^\infty  J_0 (k\rho\sin \theta_0 )
\exp (ikz\cos\theta_0 )e^{ -i\omega t}\,d\omega \:.
 \label{bes}
\end{equation}
 \begin{figure}
\includegraphics[width=0.5\textwidth]{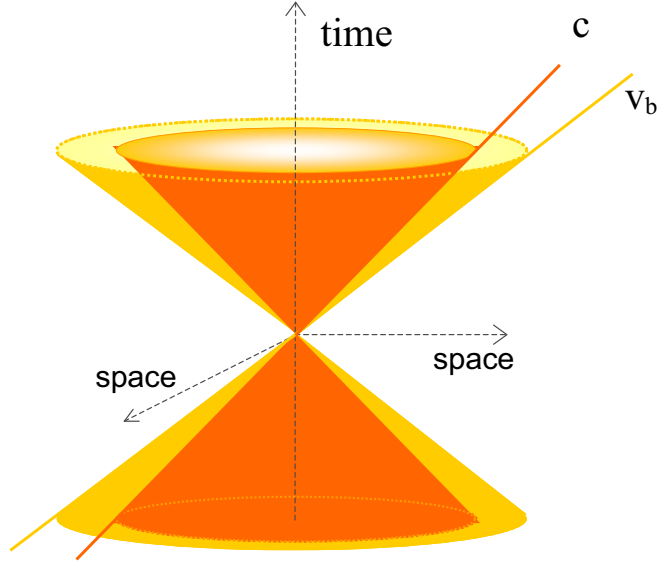}
\caption{{\small Schematic representation of the Super-Light Cone
in the Minkowski space-time (pseudo-Euclidean space). The orange
zone represents the Light-Cone, while the yellow zone around it is
the field of existence of the Bessel beam. Quantity $v_b$ is the
beam velocity for a given axicon angle, $\theta_0$. For
$\theta_0=0$, the beam is reduced to a plane wave, and its
velocity then becomes equal to $c$. In this situation,  the field
of existence of the beam goes to zero, the Super-Light Cone
narrows and becomes equal to the Light-Cone.}}
 \label{super-cone}
\end{figure}
By substituting Eq. (\ref{delta}) in Eq. (\ref{bes}) we obtain
\begin{eqnarray}
 &&U(\rho ,z,t)= \int_0^{2\pi} d\varphi \,\int_{-\infty}^\infty
 \exp \left[i\omega \left(\frac{\rho}{c} \sin\theta_0 \cos\varphi
 +\frac{z}{c}\cos\theta_0 -t\right)\right]\,d\omega \label{bes1} \nonumber \\
 && = \int_0^{2\pi} \delta \left(
\frac{\rho}{c}\sin\theta_0 \cos\varphi + \frac{z}{c}\cos\theta_0
-t\right)\,d\varphi  \,, \label{pulse}
\end{eqnarray}
where $\delta$ denotes the Dirac $\delta$-function. It can
immediately be seen  that:
\begin{description}
\item - for $\rho\leq \left|z\cot\theta_0-\frac{c}{\sin\theta_0}t\right|$
the field is zero, since the $\delta$ function has no zeros in the
integration interval;
\item - for $\rho =0\,,$  the only solution which makes the integral different from zero is
$\frac{z}{t}=v_b=\frac{c}{\cos\theta_0}$;
\item - for $\rho\geq \left|z\cot\theta_0-\frac{c}{\sin\theta_0}t\right|$, the integral
is different from zero and the field is given by
\begin{equation}
U= 4\left[\frac{\rho^2}{c^2}\sin^2\theta_0 -
\left(\frac{z}{c}\cos\theta_0- t\right)^2\right]^{-1/2}.
\end{equation}
\end{description}
The plane $z$ is therefore divided into four zones (see  Fig
\ref{cono}): in zones I and III, the field $U$ is zero, while in
zones II and IV it is different from zero. Along the straight
lines $\rho = \left|z\cot\theta_0-
\frac{c}{\sin\theta_0}t\right|$, the field is discontinuous,
remaining equal to zero on one side and going to infinity on the
other side.

Let us now analyze the field of existence of the beam in the $z-t$
plane. It is interesting to note that Eq. (\ref{pulse}) differs
from zero only if
\begin{equation}
t \leq \left| \frac{1}{c} \left( z\cos\theta_0
 +\rho\sin\theta_0\right)\right|,
  \label{tb}
\end{equation}
where $0\leq\theta_0<\theta_{max}$, and $\theta_{max} \ll \pi /2$
 depends on the experimental set-up.
Thus, the time interval in which the beam is different from zero
is
\begin{equation}
 t_{min}(\theta_0=\theta_{max}) \leq t<
 t_{max}(\theta_0=0).
 \end{equation}

Since the Bessel pulse propagates along the z-axis (see Eq.
(\ref{u})), we have that the propagation of the beam in the $z-t$
plane is within a conical surface (see Fig. \ref{beam-velocity})
similar to the Light Cone, where light velocity $c$ is replaced by
velocity $v_b = c/\cos\theta_0$, and $t$ {\it is a real quantity}:
we can say that the propagation of a Bessel pulse in the
Euclidean-space corresponds to a Super-Light Cone in the
pseudo-Euclidean space-time of Minkowski. In other words, by
introducing a second spatial coordinate, for a given value of
$\theta_0$, we obtain a Super-Light Cone like the one of Fig.
\ref{super-cone}, where straight line $v_b$, which depends on
$\theta_0$, is the beam velocity.

For $\theta_0=\theta_{max},\: v_b$ represents the border line
which determines the existence of the field: the Bessel beam
exists only in the blue zone. Inside this cone of existence, the
past Super-Light Cone, $t < 0$, represents the time interval prior
to generation of  the beam. The beam originates at $t=0$, and for
$t>0$ (future Super-Light Cone) propagates along the $z$ axis with
velocity  $v_b$ (yellow line, in Fig. \ref{super-cone}). For
$\theta_0=0$ the beam reduces to a plane wave, its velocity
becomes equal to $c$ (orange line, in Fig. \ref{super-cone}), and
the Super-Light Cone becomes the Light Cone (orange cone in Fig.
\ref{super-cone}).

Since Bessel beams have been experimentally generated and
measured, there is no doubt that they are real quantities.
Moreover, since Eq. (\ref{u}) is capable of describing the scalar
field of the beam as being due to a specific experimental set-up
\cite{mug06}, we can conclude that the Super-Light Cone places a
new upper speed limit for all objects. Massless particles can
travel not only along the Light Cone, but also along the
Super-Light Cone in the region between the Super-Cone and the
Cone, while the world-lines remain confined within the Light-Cone.
In substance, we can think that $c$ is the velocity of  light in
its simplest manifestation (wave), while more complex
electromagnetic phenomena, such as the interference among an
infinite number of waves, may originate different velocities. The
maximum value $\theta_{max}$ of  axicon angle $\theta_0$ sets the
maximum value of the beam velocity. Since the filed depth, that
is, the spatial range in which the beam exists, is proportional to
$\tan\theta_0^{-1},\: \theta_0$ can never reach the value of $\pi
/2$. If it were possible to obtain values of $\theta_0$ close to
$\pi /2$, we should have almost  immediate propagation in a
nearly-zero space, rather like an ultra fast shot destined to slow
down immediately.

The change in the upper limit of the light velocity (the Bessel
beam is ``light") does not modify the fundamental principles of
relativity and the principle of causality, as demonstrated by
recent theory dealing with new geometrical structure of space-time
\cite{car}.  The principle that  ``the speed of light is the same
for all inertial observers, regardless of the motion of the
source", remains unchanged, provided that the substitution
$c\longrightarrow v_b\, ( = c / \cos\theta_0 )$ is made in the
Lorentz transformations. In this way, the direction of the
beam-light does not depend on the motion of the source, and all
observers measure the same speed ($v_b$) in all directions,
independently of their motions.


\vspace{2 cm}

\begin{thebibliography}{99}

\bibitem{dur87}
J. Durnin, J.J. Miceli Jr., and J.H. Eberly, Phys. Rev. Lett. {\bf
58}, 1499 (1987).

\bibitem{dur87-1}
 J. Durnin, J. Opt. Soc. Am. A {\bf 4}, 651 (1987).

\bibitem{zio}
 Richard W. Ziolkowski, Phys. Rev. A {\bf 39}, 39 (1989).

\bibitem{spr}
 P. Sprangle and B. Hafizi, Phys. Rev. Lett. {\bf 66}, 837 (1991).

\bibitem{dur91}
 J. Durnin, J.J. Miceli Jr., and J.H. Eberly, Phys. Rev. Lett.
{\bf 66}, 838 (1991).

\bibitem{tan}
K. Tanaka, M. Taguchi, and T. Tanaka, J. Opt. Soc. Am. A {\bf 18},
1644 (2001).

\bibitem{rei}
 Kaido Reivelt and Peeter Saari, Phys. Rev. E {\bf 66}, 056611
 (2002).

\bibitem{saa}
P. Saari and K. Reivelt, Phys. Rev. Lett. {\bf 79}, 4135 (1997).


\bibitem{mug00}
 D. Mugnai, A. Ranfagni, and R. Ruggeri, Phys. Rev. Lett.
 {\bf 84}, 4830 (2000).

\bibitem{ale}
 I. Alexeev, K.Y. Kim, and H.M. Milchberg, Phys. Rev. Lett. {\bf
88}, 073901 (2002).

\bibitem{bes}
Ioannis M. Besieris and Amr M. Shaarawi, Optics Express {\bf 12},
3848 (2004).

\bibitem{zam}
 Miche Zamboni-Rached, Amr M. Shaarawi, and Erasmo Recami, J. Opt. Soc. Am. A {\bf 21}, 1564 (2004).

 \bibitem{saa1}
Peeter Saari and Kaido Reivelt, Phys. Rev. E {\bf 69}, 036612
(2004)

\bibitem{abr}
M. Abramowitz and Irene A. Stegun, {\it Handbook of Mathematical
Functions}, Dover Pub.,  New York, 1970, p.360,  n. 9.1.21. See
also  G.N. Watson, {\it Theory of Bessel Functions}, Cambridge,
1922.

\bibitem{bri}
L. Brillouin, {\it Wave propagation and group velocity}, Academic,
New York, 1960, p. 76, and references therein.

\bibitem{mug06} A vectorial treatment showed that the scalar field of Eq.
(\ref{u}) is that of a beam as produced by a plane wave impinging
on a ring shaped aperture placed on the focal plane of a lens (or,
more generally, of a converging system). See: D. Mugnai and I.
Mochi, Phys. Rev. E {\bf 73}, 016606 (2006). See also
http://www.arxiv.org/abs/physics/0506120.

\bibitem{car} F. Cardone and R. Mignani, {\it  Energy  and Geometry}, World Scientific
Series in Contemporary Chemical Physics - Vol. 22 (2004).




\end{thebibliography}
\end{document}